\documentclass[twocolumn,pre,aps]{revtex4-1}
\usepackage{graphics,graphicx,epsfig}
\usepackage{epsf,epstopdf,wrapfig}
\usepackage{amssymb,amsfonts,amsmath}
\usepackage{graphicx,pstricks,epsfig}
\usepackage{ifthen}
\include{epsf}

\newcommand{\beq}{\begin{equation}}
\newcommand{\eeq}{\end{equation}}
\newcommand{\bea}{\begin{eqnarray}}
\newcommand{\eea}{\end{eqnarray}}

\begin{document}

\title{Positional information, in bits}
\author{Julien O. Dubuis,$^{1,2,4}$ Ga\v{s}per Tka\v{c}ik,$^5$  Eric F. Wieschaus,$^{2,3,4}$  Thomas Gregor$^{1,2}$ and William Bialek,$^{1,2}$}
\affiliation{$^1$Joseph Henry Laboratories of Physics, $^2$Lewis--Sigler Institute for Integrative Genomics,  $^3$Department of Molecular Biology, and $^4$Howard Hughes Medical Institute, Princeton University, Princeton, New Jersey 08544 USA\\
$^5$Institute of Science and Technology Austria, Am Campus 1, A--3400 Klosterneuburg, Austria}

\begin{abstract}
Cells in a developing embryo have no direct way of ``measuring'' their physical position.  Through a variety of processes, however, the expression levels of multiple genes come to be correlated with position, and these expression levels thus form a code for ``positional information.''   We show how to measure this information, in bits, using the gap genes in the {\em Drosophila} embryo as an example.  Individual genes carry nearly two bits of information, twice as much as expected if the expression patterns consisted only of on/off domains separated by sharp boundaries.  Taken together, four gap genes carry enough information to define a cell's location with an error bar of $\sim 1\%$ along the anterior--posterior axis of the embryo.  This precision is nearly enough for each cell to have a unique identity, which is the maximum information the system can use, and is nearly constant along the length of the embryo.  We argue that this constancy is a signature of optimality in the transmission of information from primary morphogen inputs to the output of the gap gene network.
\end{abstract}

\date{\today}

\maketitle

\section{Introduction}

Building a complex, differentiated body requires that individual cells in the embryo make decisions, and ultimately adopt fates, that are appropriate to their position.    There are wildly diverging models for how cells acquire this ``positional information''  \cite{wolpert_69}, but there is a general consensus that they encode positional information in the expression levels of various key genes.  A classic example is provided by anterior--posterior patterning in the fruit fly, {\em Drosophila melanogaster}, where a small set of gap genes, and then a larger set of pair rule and segment polarity genes, are involved in the specification of the body plan \cite{nusslein+wieschaus_80}.  These genes have expression levels which vary systematically along the body axis, forming an approximate blueprint for the segmented body of the fully developed larva that we can ``read'' within hours after the start of development \cite{lawrence_92}.  

Although there is consensus that particular genes carry positional information, much less is known quantitatively about how much information is being represented.    Do the relatively broad, smooth expression profiles of the gap genes, for example,  provide enough information to specify the exact pattern of development, cell by cell, along the anterior--posterior axis?  How much information does the whole embryo actually use in making this pattern?  Answering these questions is important, in part, because we know that crucial molecules involved in the regulation of gene expression are present at low concentrations and even low absolute copy numbers, so that expression is noisy \cite{elowitz+al_02,ozbudak+al_02,blake+al_03,rosenfeld+al_05,golding+al_05,gregor+al_07b,tkacik+al_08d}, and this noise must limit the transmission of information \cite{tkacik+al_08a,ziv,deronde,inforeview}.   Is it possible, as suggested theoretically \cite{tkacik+al_08b,twb09,wtb10,twb11}, that the information transmitted through these regulatory networks is close to the physical limits set by the bounded concentrations of the different transcription factors?  To answer this and other questions, we need to measure positional information quantitatively, in bits.  We do this here using the gap genes in {\em Drosophila} as an example.

\section{Quantifying information}

\begin{figure}[bht]
\includegraphics[width =  \linewidth]{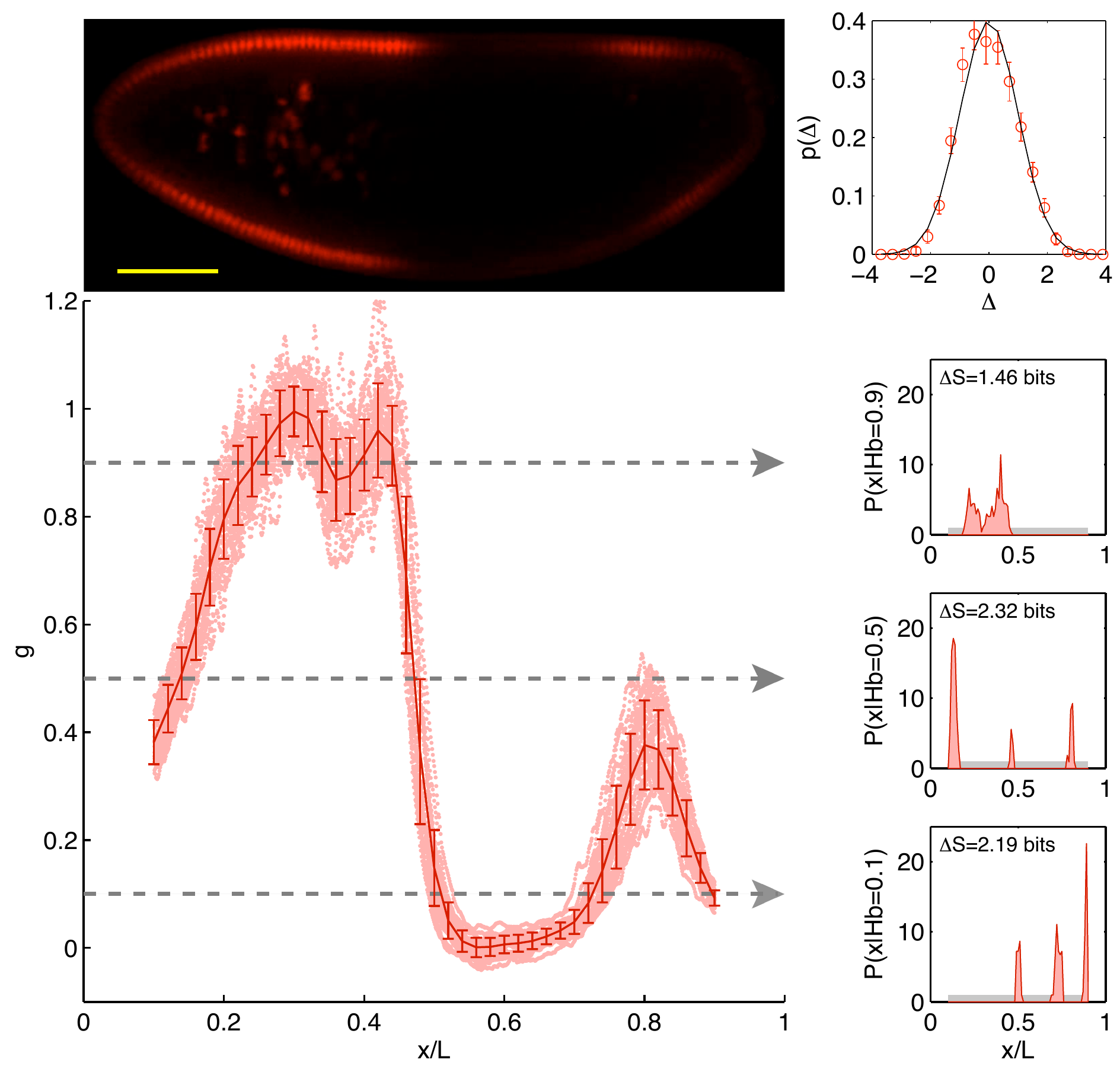}
\caption{ Positional information carried by the expression of Hunchback.
Upper left panel: optical section through the midsagittal plane of a Drosophila embryo with immunofuorescence staining against Hb protein; scale bar is $100\, \mu{\rm m}$. Lower panel left: normalized dorsal profiles of fluorescence intensity, which we identify as Hb expression level $g$, from 24 embryos (light red dots) selected in a 30 to 40 min time interval after the beginning of nuclear cycle 14 (see Methods). Means and standard deviations are plotted in darker red. Considering all points with $g = 0.1$, $0.5$, or $0.9$ yields the conditional distributions $P(x|g)$ shown at right. Note that these distributions are much more sharply concentrated than the uniform distribution $P_x(x)$, shown in light grey; correspondingly the entropies $S_{\rm f} = S[P(x|g)]$ are very much smaller than the entropy $S_{\rm i} = S[P_x(x)]$. For each $g$, we note the reduction of uncertainty in $x$ by reading out $g$, $\Delta S = S_{\rm i} - S_{\rm f}$. Upper right corner: variations in expression level around the mean at each position, estimated by the distribution of normalized relative expression, given by $\Delta = [g - \bar g (x)]/\sigma_g (x) $ (red circles with standard errors of the mean). Solid line is a zero mean/unit variance Gaussian. Details of staining, imaging, age and orientation selection, normalization and entropy estimation are given in Methods. \label{InfoHb}}
\end{figure}

Before we observe the expression levels of the relevant genes, we have no information about the position of the cell---it could be anywhere along the anterior--posterior axis of the embryo.  Mathematically this is equivalent to saying that, a priori, the position of the cell is drawn from a distribution of possibilities $P_x(x)$; in the simplest case this probability distribution is uniform, but it also is possible that cells  vary in density along the embryo's axis.  Once we observe the expression level $g$, we still don't know the precise position $x$ of the cell, but our uncertainty is greatly reduced.  In Fig \ref{InfoHb} we illustrate this idea using the gap gene {\em hunchback} ({\em hb}).  Expression levels of {\em hb} are known to vary systematically along the anterior--posterior axis of the {\em Drosophila} embryo, but we also know that expression levels can be variable across cells in the same position, both within a single embryo and across multiple embryos.  Thus, if we make a ``slice'' through the expression profile at some particular level $g$, we can't point uniquely to the position $x$ of the nucleus in which the Hunchback protein has that exact concentration.  Instead there is a range of positions which are consistent with the value of $g$, and we can summarize this range of possibilities by the conditional probability distribution, $P(x|g)$,  that a cell with expression level $g$ will be found at position $x$.  For all values of $g$ that occur in the embryo, we see that this conditional distribution is narrower or more concentrated that then nearly uniform distribution $P_x(x)$.

The probability distributions $P_x(x)$ and $P(x|g)$ provide the ingredients we need in order to make a mathematically precise version of the qualitative statement that  ``the expression level $g$ of a gene provides information about the position $x$ of the cell.''  Crucially,  the foundational result of Shannon's information theory is that there is only one way of doing this that is consistent with simple and plausible requirements, for example that independent signals should give additive information \cite{shannon_48,cover+thomas_91}. 

For any probability distribution we can define an entropy $S$, which is the same quantity that appears in statistical mechanics and thermodynamics; for the two distributions here we have
\begin{eqnarray}
S[P_x(x)] &=& - \int dx\, P_x(x) \log_2 [P_x(x)]\, {\rm bits},\\
S[P(x|g)] &=& - \int dx\, P(x|g) \log_2 [P(x|g)]\, {\rm bits}.
\end{eqnarray}
For example, if we measure $x$ from $0$ to $L$ along the length of embryo, then a uniform distribution of cells corresponds to $P_x(x) =1/L$, and this has the maximum possible entropy $S[P_x(x)]$.   The intuition that the conditional distribution   $P(x|g)$ is narrower or more concentrated  than $P_x(x)$ is quantified by the fact that the entropy $S[P(x|g)] $ is smaller than $S[P_x(x)]$, and this reduction in entropy is exactly the information that observing $g$ provides about $x$, here measured in bits.  As an example, if observing the expression level $g$ tells us, with complete certainty, that the cell is located in a small region of size $\Delta x$, then the gain in information is $I \equiv S[P_x(x)]  - S[P(x|g)] = \log_2 (L/\Delta x)\,{\rm bits}$.

If we choose a cell at random, we will see an expression level $g$ drawn from the distribution $P_g(g)$.  The average information that this expression level provides about position is then
\begin{eqnarray}
I_{g\rightarrow x} &=& \int dg\,P_g(g) \left( S[P_x(x)] - S[P(x|g)]  \right) ,\label{I=diffS}\\
&=& \int dg \int dx\, P(g,x) \log_2 \left[{{P(g,x)}\over{P_g(g) P_x(x)}}\right],
\label{MIdef}
\end{eqnarray}
where $P(g,x)$ is the joint probability of observing a cell at $x$ with expression level $g$, and we have rearranged the terms to emphasize the symmetry---information which the expression level provides about the position of the cell is, on average, the same as the information that the position of the cell provides about the expression level.  This average information is called the mutual information between $g$ and $x$.  Again we emphasize that this measure of information is {\em not} one among many equally good possibilities, it is unique.

Because information is mutual, we can also write $I_{g\rightarrow x}$ in terms of the distribution of expression levels $g$ that we find in cells at a particular position, $P(g|x)$,
\begin{equation}
I_{g\rightarrow x} = \int dx\,P_x(x) \left( S[P_g(g)] - S[P(g|x)]  \right) .\label{Idirect}
\end{equation}
This emphasizes that the amount of information that can be conveyed is limited both by the overall dynamic range of expression levels, which determines $S[P_g(g)]$, and by the variability or noise in expression levels at a fixed position, which is measured by $S[P(g|x)]$. It will be useful that the distribution of expression levels at one point, $P(g|x)$, is approximately Gaussian, as shown at the upper right in Fig \ref{InfoHb}.

In what follows we will use Eq (\ref{Idirect}) to make a ``direct'' measurement of information, while Eq (\ref{I=diffS}) invites to try and ``decode'' the information carried by the expression levels to recover estimates of the position $x$ of each cell.  Each approach has a natural generalization to the case where information is conveyed not by the expression level of one gene but by the combined expression levels of multiple genes $\{g_{\rm i}\}$, and we will explore this as well.  It is important to emphasize that the number of bits of information carried by the gene expression levels has meaning independent of the mechanisms by which this coding is established.  Thus, at one extreme, it could be that each cell sets its expression levels independently in response to some primary morphogen (such as Bicoid in the {\em Drosophila} embryo \cite{driever+vollhard_88a,driever+vollhard_88b,bcd_review}), while at the other extreme the spatial patterns of expression could arise entirely from communication between neighboring cells, in a Turing--like mechanism \cite{turing_52,more_turing}.  In these different extremes, the precise value of the positional information places different quantitative constraints on the underlying mechanisms, but in all cases the number of available bits tells us about the reliability and complexity of the pattern that can be constructed from the local expression levels alone.

\section{Information carried by single gap genes}

Estimating the mutual information that one gene expression level provides about position requires, from Eq (\ref{Idirect}), that we obtain a good estimate of the conditional distribution $P(g|x)$.  Using immunofluorescent staining, we can measure $g$ vs. $x$ along the anterior--posterior axis of single {\em Drosophila} embryos, and by making such measurements on multiple embryos, as shown in Fig \ref{InfoHb}, we obtain many samples of the expression level at corresponding positions, and then from these samples we can build up an estimate of the  distribution $P(g|x)$.  Because expression profiles vary systematically with time during nuclear cycle 14, it is important to make these measurements on embryos in a limited time class, which,  we do by taking the length of the cellularization membrane as  a proxy for time (\cite{myasnikova}; see Methods).  We also confine our attention to the central 80\% of the anterior--posterior axis, both because quantitative imaging at the poles is more difficult and because we know that there are additional genes associated specifically with terminal patterning.

As has been addressed in other contexts (see Methods), care is required to be sure that the finite number of samples we collect is sufficient  to get a reliable estimate of $I(g;x)$, but once we have control over the potential systematic errors the statistical errors in our measurements are very small.  Analysis of the data in Fig \ref{InfoHb} shows that the expression level of Hunchback provides $I_{g_{\rm Hb}\rightarrow x} = 2.26\pm 0.04\,{\rm bits}$  of information about the position of a cell along the middle 80\% of the anterior--posterior axis.   We can repeat this analysis for  the gap genes {\em kr\"uppel}, {\em giant} and {\em knirps}, in addition to {\em hunchback}, and we find
 $I_{g_{\rm Kr}\rightarrow x} = 1.95\pm 0.07\,{\rm bits}$,
 $I_{g_{\rm Gt}\rightarrow x} = 1.84\pm 0.05\,{\rm bits}$, and
 $I_{g_{\rm Kni}\rightarrow x} = 1.75\pm 0.05\,{\rm bits}$.

In all cases, the expression of a single gene carries much more than one bit of information, indeed more nearly two bits.  The conventional view of the gap genes is that they are characterized by domains of expression, with boundaries, and the sharpness of the boundary often is taken as a measure of precision.  But if the patterns of expression were perfect on/off domains with infinitely sharp boundaries, then the expression level could provide {\em at most} one bit of information about position.  Our result that gap genes provide nearly two bits of information about position demonstrates that intermediate expression levels are sufficiently reproducible from embryo to embryo that they carry significant amounts of positional information, and that the view of domains and boundaries misses almost half of this information.

\section{How much information does the embryo use?}

\begin{figure}[bt]
\includegraphics[width =  \linewidth]{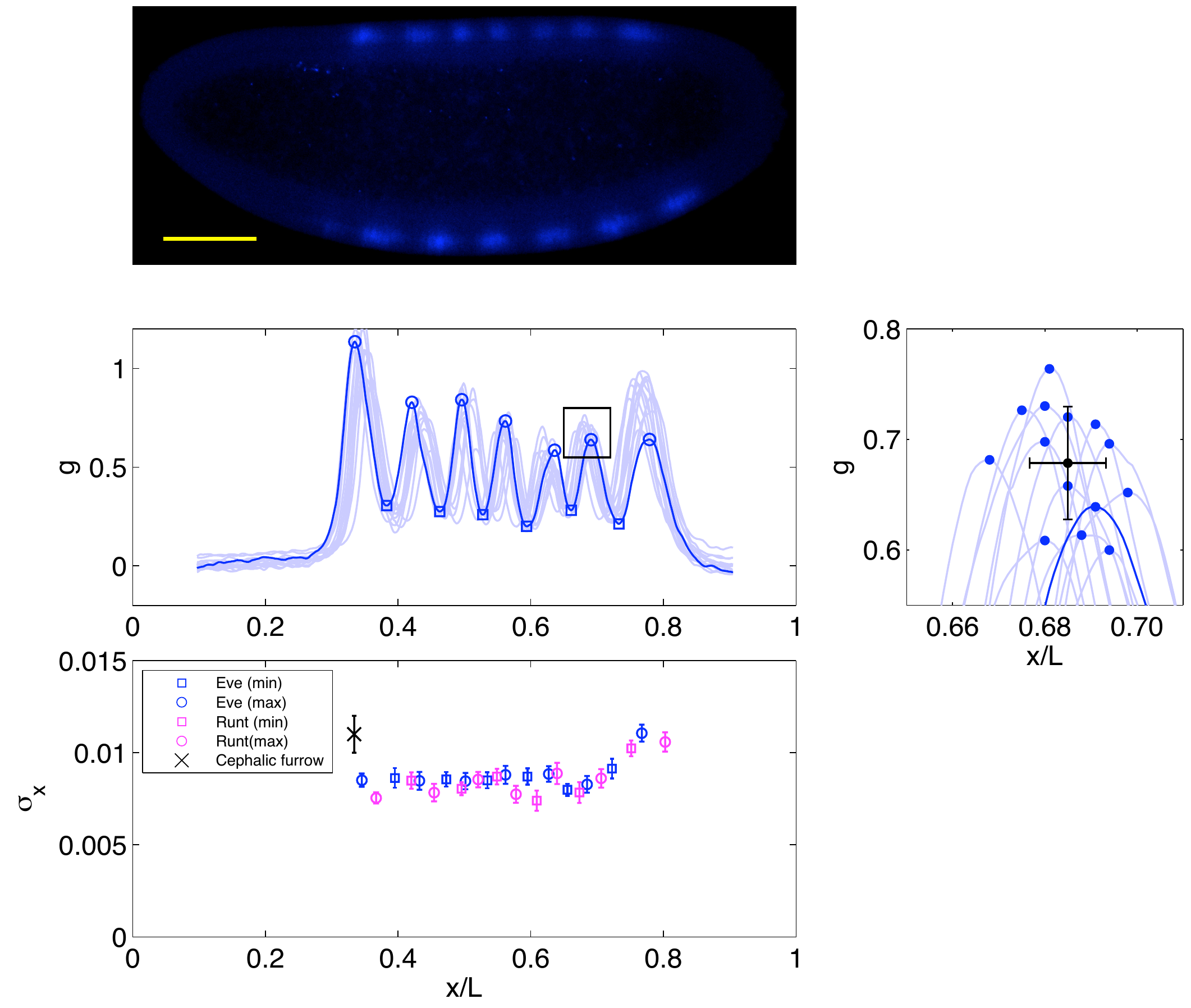}
\caption{Reproducibility of multiple pattern elements along the anteriorÐposterior axis.
Top panel: optical section through the midsagittal plane of a {\em Drosophila} embryo with immunofuorescence staining against Eve protein; scale bar is $100\, \mu{\rm m}$. Middle panels: normalized dorsal profiles of fluorescence intensity from 12 embryos selected in a 40 to 50 min time window after the beginning of nuclear cycle 14 (light blue lines); dorsal profile of top panel embryo is in darker blue. Zooming in on a single peak (as shown at right), we can measure the standard deviation of both the expression level and position of this element in the pattern. Bottom panel summarizes results from such measurements on Even--skipped (blue) and Runt (magenta), plotting the standard deviation of the position $\sigma_x$ as a function of the mean position $\bar{x}$, together with a similar measurement on the reproducibility of the cephalic furrow. Note that all of the elements are positioned with 1\% accuracy or better.
\label{pair_rule}}
\end{figure}

If the expression profile of each gap gene were described by on/off domains with sharp boundaries, not only would a single gene carry at most one bit of information, four genes taken together could carry at most four bits---and this would happen only if the spatial arrangement of the different expression domains were carefully aligned to minimize redundancy.  Four bits of information corresponds to, at most, 16 reliably distinguishable states encoded by these genes, which  seems small compared with the complexity of the pattern that eventually forms. But how much information does the embryo really need, or use?  At best, every nucleus could be labelled with a unique identity, so that with $N$ nuclei the embryo could make use of  $\log_2 N$ bits.  Along the anterior--posterior axis, we can count nuclei in a single mid--sagittal slice through the embryo, and in the middle $80\%$ of the embryo where the images are clearest we have  $N= 58\pm 4$ along the dorsal side and $N= 59\pm 4$ along the ventral side, where the error bars represent standard deviations across a population of $57$ embryos  in nuclear cycle 14, corresponding to $5.9\pm 0.1$ bits of information.   But do individual cells in fact ``know'' their identity?  More precisely, are the elements of the anterior--posterior pattern specified with single cell resolution?

There are several experiments suggesting that elements of the final body plan of the maggot can be traced to identifiable rows of cells along the anterior--posterior axis \cite{gergen}, which is consistent with the idea that each row of cells has a reproducible identity.  More quantitatively, we can ask about the reproducibility of various pattern elements in early development, elements that appear not long after the expression patterns of the gap genes are established.  A classic case is the cephalic furrow, which can be observed in live embryos and is known to have a position along the anterior--posterior axis that is reproducible with $\sim 1\%$ accuracy (see, for example, Ref \cite{gregor+al_07b}).

Is the cephalic furrow special, or can the embryo more generally position pattern elements with $\sim 1\%$ accuracy?  The striped patterns of pair rule gene expression allow us to ask about the position of multiple pattern elements, seven peaks and six troughs of expression along the anterior--posterior axis.  As shown in Fig \ref{pair_rule}, all of these elements have positions that are reproducible to within $1\%$ of the embryo length.  This strongly suggests that all cells ``know'' their position along the anterior--posterior axis with $\sim 1\%$ precision.

\section{Decoding the positional information carried by multiple genes}

Do the four gap genes, taken together, carry enough information to specify position with $\sim 1\%$ accuracy?  To answer this, it is useful to look more directly at how the information is encoded.  We observe the expression levels $g_{\rm i}$, with ${\rm i} = 1,\, 2,\, 3,\, 4$.  At each point $x$ there are average values of these expression levels $\bar g_{\rm i}(x)$, and there are fluctuations $\delta g_{\rm i}$.  Let us assume that these fluctuations have a Gaussian distribution.  If we look just at one gene, this means that the statistics of the fluctuations are described completely by the mean and the variance $\sigma_{\rm i}^2 (x)$, so that if we look at the same position $x$ in many embryos we will see a distribution of expression levels
\begin{equation}
P(g_{\rm i}|x) = {1\over\sqrt{2\pi\sigma_{\rm i}^2}} \exp\left[ - {{(g - \bar g_{\rm i}(x))^2}\over{2\sigma_{\rm i}^2 (x)}}\right] ,
\end{equation}
and this is in good agreement with the measurements in Fig \ref{InfoHb}.  If we look at many genes simultaneously, we have not just the variances of each gene but also the correlations or covariances among the genes, which define a matrix $C_{\rm ij} (x)$.  The joint distribution of expression levels at one point is then
\begin{widetext}
\begin{equation}
P(\{g_{\rm i}\}|x) = {1\over\sqrt{(2\pi)^4 \det C}} \exp\left[ -{1\over 2}\sum_{{\rm i},{\rm j}=1}^4 
(g_{\rm i} - \bar g_{\rm i}(x)) \left( C^{-1}\right)_{\rm ij} (g_{\rm j} - \bar g_{\rm j}(x)) \right],
\label{jointPg}
\end{equation}
\end{widetext}
where $C^{-1}$ denotes the inverse of the matrix $C$ and $\det C$ denotes its determinant.  But to ``read'' the information carried by the expression levels, we need to ask for the distribution of positions that are consistent with a particular set of expression levels that we might observe.  By Bayes' rule, this can be written as
\begin{equation}
P(x| \{g_{\rm i}\}) = {{P(\{g_{\rm i}\}|x)P_x(x)}\over{P_g (\{g_{\rm i}\})}} ,
\end{equation}
where $P_x(x)$ is, as before, the (nearly uniform) distribution of cell positions and $P_g (\{g_{\rm i}\})$ is the (joint) distribution of expression levels averaged over all cells in the embryo.

\begin{figure*}
\includegraphics[width =  0.9\linewidth]{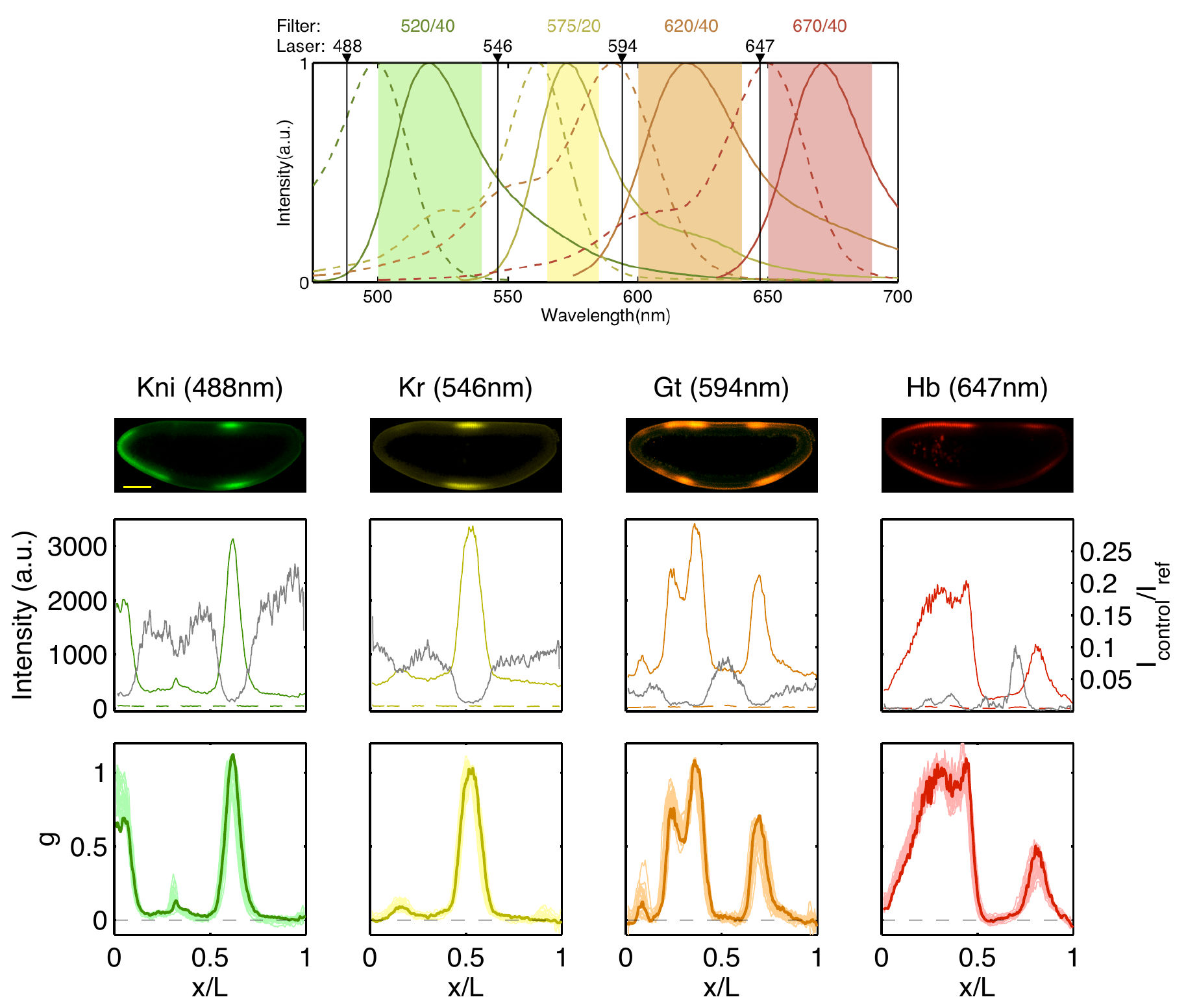}
\caption{Simultaneous immunostaining of Hunchback, Kruppel, Giant and Knirps.
Top panel: absorption (dashed lines) and emission (plain lines) spectra of the secondary dies used for simultaneous immunostaining of 4 proteins, the laser excitation wavelength (in black) and the position and bandwidth of the filters for each detection channel.  Below are optical sections through the midsagittal plane of a single {\em Drosophila} embryo with co--immunofuorescence staining against Knirps (green), Kruppel (yellow), Giant (orange) and Hunchback (red); scalebar is $100\, \mu{\rm m}$. To estimate the crosstalk for each channel, we compare the intensity profile in the sample embryo (plain line) with a control embryo of similar age and orientation for which all of the channels but the considered one have been co-stained (dashed line). The ratio between the two curves is plotted in grey; note the different scale at right. The bottom panels show the dorsal expression levels of the four gap genes for 24 embryos (light colors). The dorsal expression levels of the sample embryo are plotted in darker colors.
\label{fig:4color}}
\end{figure*}

If the noise levels are small, then $P(x| \{g_{\rm i}\}) $ will be sharply peaked at some $x_*(\{g_{\rm i}\})$ which is our best estimate of the position given our observations on the expression levels.  Expanding around this estimate, we find that the distribution is approximately Gaussian,
\begin{equation}
P(x| \{g_{\rm i}\}) \approx {1\over\sqrt{2\pi\sigma_x^2}} \exp\left[ - {{(x-x_*(\{g_{\rm i}\}))^2}\over{2\sigma_x^2}}\right] ,
\label{xgauss}
\end{equation}
where the error in our position estimate is defined by
\begin{equation}
{1\over {\sigma_x^2}} = \sum_{{\rm i},{\rm j}=1}^4 \left[ {{d\bar g_{\rm i}(x)}\over{dx}}  \left( C^{-1}\right)_{\rm ij}{{d\bar g_{\rm i}(x)}\over{dx}}\right] {\Bigg |}_{x = x_*(\{g_{\rm j}\})} .
\label{sigmax_def}
\end{equation}

\begin{figure}[t]
\includegraphics[width =  \linewidth]{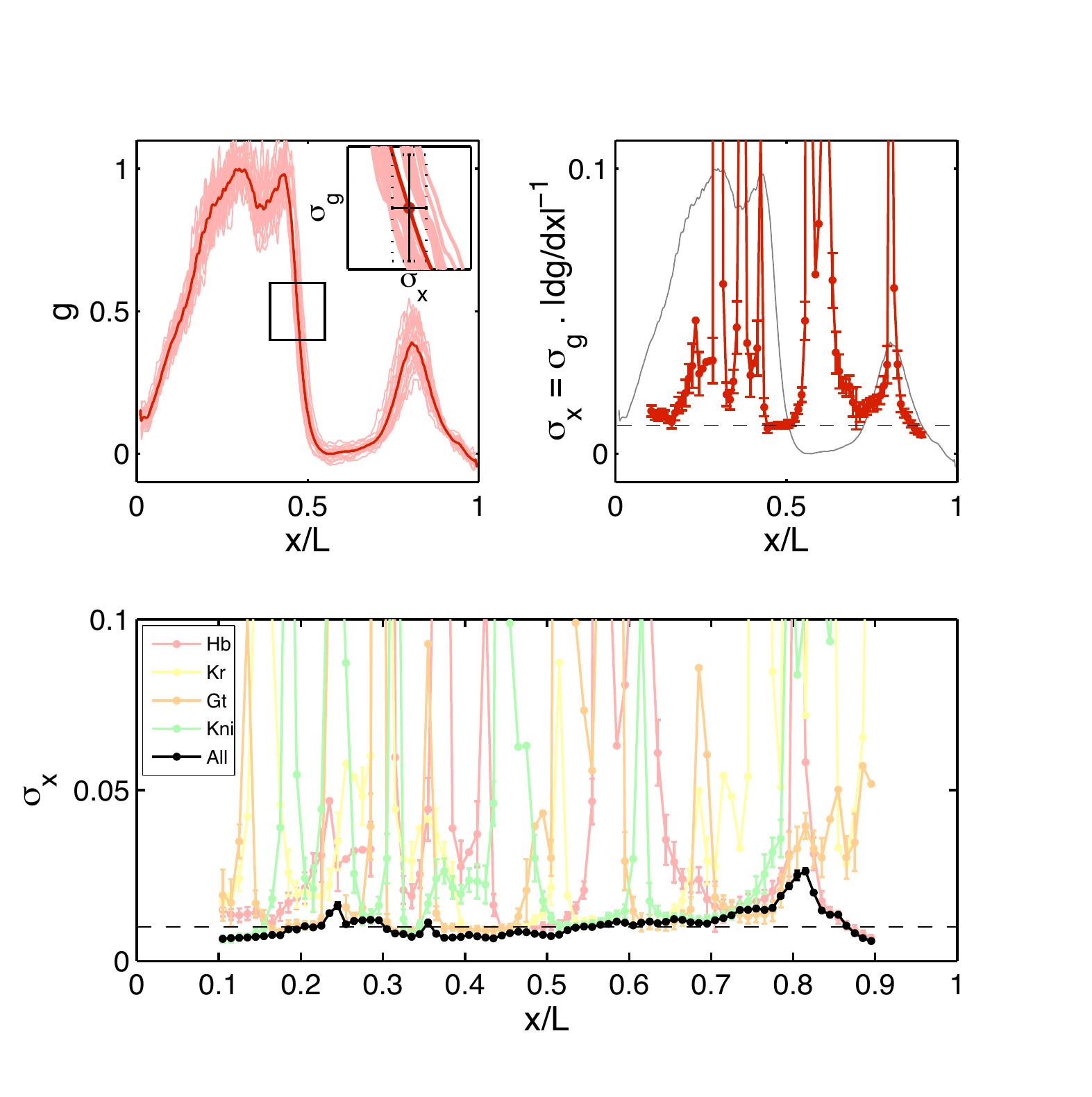}
\caption{Positional error as a function of position.
Upper left panel: geometrical interpretation of the positional error of a single gene at a given position. $\sigma_x(x)$ is proportional to the reproducibility of the profiles and is inversely proportional to the derivative of the mean profile. The upper right panel summarizes the error in positional estimates based on the Hunchback level for a hundred points along the anteroposterior axis. Lower panel: error in positional estimates based on the expression levels of the four gap genes together (in black), from Eq (\ref{sigmax_def});  error bars are from bootstrapping. For reference, positional errors based on individual expression levels are plotted in lighter colors in the background. Note that the net positional error is nearly constant and equal to 1\% of the total egg length.
\label{fig:sigma_x}}
\end{figure}

Equation (\ref{sigmax_def}) tells us the precision with which expression levels encode position:  observing the expression levels $\{g_{\rm i}\}$ allows us (or the cell!) to specify position with an ``error bar'' $\sigma_x$.  Note that this error could be different at different points in the embryo, so really we should write $\sigma_x(x)$.  
Checking our intuition, we see  that this error bar is smaller when the variability in expression is smaller (smaller $C$), when the mean spatial variations in expression levels are stronger (larger $d\bar g_{\rm i}/dx$), or when we can sum over more information carried by more genes.  We can define a similar quantity based on measurements of a single gene,
\begin{equation}
{1\over {\sigma_x(x)}} =  {\bigg |}{{d\bar g_{\rm i}(x)}\over{dx}}{\bigg |} {1\over{\sigma_{\rm i}(x)}} ,  \label{sigmax_onegene}
\end{equation}
and this construction is shown schematically at the top of Fig \ref{fig:sigma_x}.  Note that when $\sigma_x$ is small, we can justify our approximation that $P(x|\{g_{\rm i}\})$ is sharply peaked, but when $\sigma_x$ becomes large it is more rigorous simply to say that we don't have much information about $x$, rather than trying to give a more quantitative interpretation.

Importantly, all the terms in Eq (\ref{sigmax_def}) are experimentally accessible.  Measurements of the average expression profiles $\bar g_{\rm i}(x)$ are standard.  Ideally, to measure the covariance matrix $C_{\rm ij}(x)$ we should observe all four genes at once, in multiple embryos, and such experiments are shown in Fig \ref{fig:4color}.  Alternatively, one can make measurements in which pairs of genes ${\rm i,j}$ are stained, and each such experiment contributes to estimates of the matrix elements $C_{\rm ii}$, $C_{\rm jj}$, and $C_{\rm ij}$.  With care, as described in Methods, such pairwise experiments can be merged to give the same results as the more direct quadruple staining.  The major difficulty in the quadruple staining experiment is to avoid spectral crosstalk among the different fluorescence signals, but as noted in the Methods modest amounts of crosstalk actually don't change our estimate of $\sigma_x$.

Measurements of $\sigma_x$ are summarized in Fig \ref{fig:sigma_x}.  Remarkably, the reliability of position estimates based on the four gap genes is $\sim 1\%$, almost precisely equal to the observed reproducibility with which pattern elements are positioned along the anterior--posterior axis.  This is strong evidence that the gap genes, taken together, carry the information needed to specify the full pattern.  Further, this positional accuracy is almost constant along the length of the embryo, which again is consistent with what we see in Fig \ref{pair_rule}.  This constancy emerges in a nontrivial way from the expression profiles, the  noise levels, and the correlation structure of the noise. If we try to make estimates based on one gene, we can reach $\sim 1\%$ accuracy in a very limited region of the embryo, and estimates from the different genes have their optimal precision in different places.  The detailed structure of the spatial profiles insures that these signals can be combined to give nearly constant accuracy. 

If the errors in estimating position really are Gaussian, as in Eq (\ref{xgauss}), then we can substitute into Eq (\ref{I=diffS}) to show that $I = \langle \log_2 [L/(\sigma_x \sqrt{2\pi e})]\rangle$, where $L$ is the length of the embryo and $\langle \cdots \rangle$ denotes an average over the possible position dependence of the error $\sigma_x$.  Computing this average, we have $I =  4.57\pm 0.02\,{\rm bits}$.  On the other hand, we can use the distribution of expression levels at each position, Eq (\ref{jointPg}), to compute the information directly as in Eq (\ref{Idirect}), and we find $I=4.97\pm 0.23\,{\rm bits}$.  The agreement between these estimates supports our approximations, and gives us confidence that the measurement of $\sigma_x$ if Fig \ref{fig:sigma_x} really does characterize the encoding of positional information by the gap genes.

\section{A signature of optimization?}

The discussion thus far concerns the amount of information that actually is transmitted by the levels of gap gene expression.  But we know that  the capacity to transmit information is strictly limited by the available numbers of molecules, and that significant increases in information capacity would require vastly more than proportional increases in these numbers \cite{tkacik+al_08a}.  Given these limitations, however, cells can still make more or less efficient use of the available capacity.  To maximize efficiency, the input/output relations and noise characteristics of the regulatory network must be matched to the distribution of input transcription factor concentrations \cite{tkacik+al_08b}.  This matching principle has a long history in the analysis of neural coding \cite{barlow_59,laughlin_81,brenner+al_00}, and in Ref \cite{tkacik+al_08b} it was suggested that the regulation of Hunchback by Bicoid might provide an example of this principle.  Here we consider the generalization of this argument to the gap gene network as a whole.

If we imagine that there is a single primary morphogen,  then the expression levels of the different gap genes, taken together, can be thought of as encoding the concentration $c$ of this morphogen.   By analogy with Eq (\ref{sigmax_def}), these expression levels can be decoded with some accuracy $\sigma_c^{\rm eff}(c)$, which itself depends on the mean local concentration.  The key result of Ref \cite{tkacik+al_08b} is that, when noise levels are small, all the ``symbols'' in the code should be used in proportion to their reliability, or in inverse proportion to their variability.  Thus, if we point to a cell at random, we should see that the concentration of the primary morphogen is drawn from a distribution
\begin{equation}
P_{\rm input}(c) = {1\over {Z}} \cdot {1\over {\sigma_c^{\rm eff} (c)}},
\label{Popt2}
\end{equation}
where the constant $Z$ is chosen to normalize the distribution.  But the input is a morphogen, so its variation is connected with the physical position $x$ of cells along the embryo: we should have $c = c(x)$.  Then if the cells are distributed uniformly along the length of the embryo, the probability that we find a cell at $x$ is just $P(x) = 1/L$, and hence we must have
\begin{eqnarray}
P_{\rm input}(c) dc &=& P(x) dx = {{dx}\over L} \\
\Rightarrow P_{\rm input}(c)  &=& {1\over L} {\bigg |}{{dc(x)}\over {dx}} {\bigg |}^{-1} .
\label{P_jacob}
\end{eqnarray}

We have two expressions for the distribution of input transcription factor concentrations:  Eq (\ref{P_jacob}), which expresses the role of the input as morphogen, encoding position $x$, and Eq (\ref{Popt2}), which expresses the solution to the problem of optimizing information transmission through the network of genes that respond to the input.  Putting these expressions together, we have
\begin{equation}
{{Z}\over L}= {1\over {\sigma_c^{\rm eff} (c)}}  {\bigg |}{{dc(x)}\over {dx}} {\bigg |} =  \sigma_x (x),
\label{opt_result}
\end{equation}
where in the last step we recognize the equivalent positional noise  $\sigma_x (x)$ by analogy with Eq (\ref{sigmax_def}).  Thus, optimizing information transmission predicts that the positional uncertainty $\sigma_x (x)$ will be constant along the length of the embryo, as observed in Fig \ref{fig:sigma_x}.   A more detailed version of this argument is given in the Appendix.

\section{Discussion}

The final result of embryonic development appears precise and reproducible.  Less is known quantitatively about the degree of this precision, and about the time at which precision first becomes apparent.  Our central result is that, in the early {\em Drosophila} embryo, the patterns of gap gene expression provide enough information to specify the positions of individual cells with a precision of $\sim 1\%$ along the anterior--posterior axis.    This is the same precision with which subsequent pattern elements are specified,  from the pair rule expression stripes through the cephalic furrow, so that all the required information is available from a local readout of the gap genes.

The precise value of the information that we observe is also interesting.  It corresponds to being able to locate any nucleus with an error bar that is smaller than the distance to its neighbor, but the total number of bits is not quite large enough to specify the position of every cell uniquely.  The difference is that when we make an estimate with error bars, the estimate comes from a distribution with tails, and the (small) overlap of the tails of these distributions means that one cannot quite identify every cell.  It is possible that cells in fact do not quite have unique identities, or that the missing information is hiding in correlations among the errors at different points:  although the gap genes encode position with an error bar, the difference between  positions coded by expression levels in neighboring cells could have a much smaller error bar.  While further experiments are required to settle this issue, we find it remarkable that the gap gene expression levels carry so much information, so that an enormously precise pattern is available very early in development. 

The information that gene expression levels can carry about position is limited by noise.  In particular, both because the concentrations of transcription factors are low, and because the absolute copy numbers of the output proteins are small, there are physical sources of noise that cannot be reduced without the embryo investing more resources in making these molecules.  Given these limits, it still is possible to transmit more information through the gap gene network by ``matching'' the distribution of input signals to the noise characteristics of the network.  Although this matching condition is in general complicated, in the limits that the noise is small it can be expressed very simply:  the density of cells along the anterior--posterior axis should by inversely proportional to the precision with which we can infer position by decoding the signals carried n the gap gene expression levels.  Since cells are almost uniformly distributed at this stage of development, this predicts that an optimal network would have a uniform precision, and this is what we find.  This uniformity emerges despite the complex spatial dependence of all the ingredients, and thus seems likely to be a signature of selection for optimal information transmission.

\begin{acknowledgments}
This work was supported in part by NSF Grants PHY--0957573 and CCF--0939370, by NIH Grants P50 GM071508, R01 GM077599 and R01GM097275, by the Howard Hughes Medical Institute,  by the WM Keck Foundation, and by Searle Scholar Award 10--SSP--274 to TG.  
\end{acknowledgments}

\section*{Appendix}

Here we give a detailed version of the arguments leading to Eq (\ref{opt_result}).   Consider the case where information flows from a single input transcription factor (such as Bicoid) to a set of $K$ output genes (the gap genes).  The concentration of the input is $c$, and the output genes have expression levels $g_1, \, g_2 , \, \cdots , \, g_K$ \cite{twb09,wtb10,twb11}.  Different cells in the embryo experience different values of $c$, depending on their position, and if we choose a cell at random it sees a concentration drawn from the distribution $P_{\rm in}(c)$.  The network responds to this input, generating expression levels that are drawn from the distribution $P(\{g_{\rm i}\}| c)$; it will also be useful to define the (joint) distribution of output expression levels,
\begin{equation}
P_{\rm out}(\{ g_{\rm i}\}) =  \int dc\, P_{\rm in}(c)  P(\{g_{\rm i}\}| c) .
\end{equation}
The information that flows from input to output can then be written as the difference of entropies, as in Eq (\ref{I=diffS}),
\begin{widetext}
\begin{equation}
I(\{g_{\rm i}\}; c) = -  \int dc\, P_{\rm in}(c) \log_2 P_{\rm in}(c)   - \int d^K g \, P_{\rm out}(\{ g_{\rm i}\}) S[P(c| \{ g_{\rm i}\})] ,
\label{ent_diff}
\end{equation}
\end{widetext}
where, from Bayes' rule, we have
\begin{equation}
P(c|\{ g_{\rm i}\}) = {{ P(\{g_{\rm i}\}| c) P_{\rm in}(c)}\over{P_{\rm out}(\{ g_{\rm i}\})}} .
\end{equation}

The transmitted information $I(\{g_{\rm i}\}; c)$ depends both on the characteristics of the gene network, expressed in $P(\{g_{\rm i}\}| c)$, and on the distribution of input signals, $P_{\rm in}(c)$.  In particular, the irreducible noise associated with the finite number of available molecules is encoded by the details of $P(\{g_{\rm i}\}| c)$.  Given these constraints it still is possible to maximize information transmission by proper choice of the input distribution \cite{shannon_48,cover+thomas_91}.  In general this optimization is a hard problem, but we can make progress if we assume that the noise is small, and we will argue that this is a good approximation.  

In Eq (\ref{ent_diff}), we need to take an average over the full distribution of output expression levels, $P_{\rm out}(\{ g_{\rm i}\})$,  This distribution is broadened by two effects.  First, the inputs $c$ are varying, and  the outputs would vary in response.  Second, even when the input $c$ is fixed, the outputs $\{g_{\rm i}\}$ may vary because of noise.  We will assume that noise is small in the sense that the first effect is much larger than the second, so that we can average over outputs by assuming that the output is always equal to its average value, $g_{\rm i} = \bar g_{\rm i}(c)$, and then averaging over the input $c$.  In this approximation,  the information becomes
\begin{widetext}
\begin{equation}
I = -  \int dc\, P_{\rm in}(c) \log_2 P_{\rm in}(c) - \int dc\,P_{\rm in}(c) S_{\rm cond}^{(c)}(\{ g_{\rm i} = \bar g_{\rm i}(c)\}) , 
\end{equation}
\end{widetext}
where $ S_{\rm cond}^{(c)}( \{ g_{\rm i}\}) = S[P(c|\{g_{\rm i}\})]$.  To find the distribution of inputs that maximizes the information, we introduce as usual a Lagrange multiplier to fix the normalization of $P_{\rm in}(c)$ and solve
\begin{equation}
{{\delta}\over{\delta P_{\rm in}(c)}} \left[ I - \Lambda \int dc \, P_{\rm in}(c) \right] = 0.
\end{equation}
The result is
\begin{equation}
P_{\rm in}(c) = {1\over Z} \exp\left[ - (\ln 2) S_{\rm cond}^{(c)}(( \{ g_{\rm i} = \bar g_{\rm i}(c)\}) \right],
\label{Popt}
\end{equation}
where  is $Z$ chosen to normalize the distribution.  The only approximation we have made thus far is to assume that the noise is small.  But if the noise is  also approximately Gaussian---given knowledge of the gene expression levels $\{ g_{\rm i} \}$, we know the input concentration to within some error bar $\sigma_c^{\rm eff} (c)$, which itself depends on the actual value of the input---then $S_{\rm cond}^{(c)} = \log_2 [\sqrt{2\pi e }\sigma_c^{\rm eff} (c)]$, and 
\begin{equation}
P_{\rm in}(c) = {1\over {Z}} \cdot {1\over {\sigma_c^{\rm eff} (c)}},
\label{Popt2a}
\end{equation}
corresponding to Eq (\ref{Popt2}) in the text.  As discussed in Ref \cite{tkacik+al_08b}, this tells us that the system can optimize information transmission by using the ``symbols" $c$ in proportion to their reliability.  

Notice that the size of the noise in the system can be summarized by $\sigma_x$ itself.  Not only do we find, experimentally, that this is nearly constant, it is also very small, and in particular smaller than the distances over which the output of any single gap gene varies significantly.  Thus, in retrospect, the effective noise really is small, as assumed above, which justifies the approximation leading to Eq (\ref{Popt}).   This derivation can be generalized to cases where there are multiple independent morphogen inputs, each varying along $x$.

\section*{Methods}

{\em Fixation and  staining.} All embryos were collected at 25 C and dechorionated in 100\% bleach for 2 minutes, then heat fixed in a saline solution (NaCl,Triton X-100) and vortexed in a vial containing 5 mL of Heptane and 5 mL of methanol for one minute. They were then rinsed and stored in methanol at -20 C. Embryos were labeled with fluorescent probes. We used rat anti--Kni, guinea pig anti--Gt, rabbit anti--Kr (gift of C. Rushlow), and mouse anti--Hb. Secondary antibodies were respectively conjugated with Alexa-488 (rat), Alexa-568 (rabbit), Alexa-594  (guinea pig) and Alexa-647 (mouse) from Invitrogen. Embryos were mounted in AquaPolymount from Polysciences, Inc.

{\em Imaging and profile extraction.}  All embryos where imaged on a Leica SP5 laser-scanning confocal microscope and image analysis routines were implemented in Matlab software (MATLAB, MathWorks, Natick, MA). Images were taken with a Leica $20\times$ HC PL APO NA 0.7 oil immersion objective, and sequential excitation wavelengths of 488, 546, 594 and 633 nm. For each embryo, three high--resolution images ($1024\times 1024$ pixels, with 12 bits and at 100Hz) were taken along the anteroposterior axis (focused at the midsagittal plane) at $1.7\times$ magnified zoom and three times frame-averaged. With these settings, the linear pixel dimension corresponds to $0.44 \pm 0.01\,\mu{\rm m}$. Profiles were extracted by sliding, in software, a disk of the size of a nucleus along the edge of the embryo in the midsagittal plane and computing the average intensity of its pixels. The coordinates of the disk centers were projected on the anterior--posterior and  dorso-ventral axes of the embryo. Two curves, corresponding to the dorsal and ventral sides of the embryo, were constituted. For consistency, only dorsal profiles are used in our analysis.    We follow the methods of Ref \cite{gregor+al_07b} to convert measured fluorescence intensities into normalized protein concentrations.

{\em Determining the age of the embryo.}   The time since initiation of nuclear cycle 14 was determined by the length of the dorsal cellularization membrane \cite{merrill+al_88,myasnikova}.  A series of $N=10$ brightfield movies of wildtype OreR embryos was used to obtain a calibration of cellularization progression. The measured length of each immunostained embryo used in our analysis was compared with the reference to convert length into time. The errorbar in age estimation by this method is $\pm 3\,{\rm min}$. Embryos were sorted according to five time intervals (0-10 mins, 10-30 mins, 30-40 mins, 40-50 mins and 50-60 mins), and our analysis here focuses on the 30-40 min class.


{\em Information in single genes.}  Measurements on the expression profiles of a single gene in multiple embryos can be thought of as providing many samples out of the joint distribution $P(g,x)$.  To compute the mutual information between $g$ and $x$, we discretize the two continuous axes into a number of bins; along the $g$ axis we use these bins adaptively, so that the histogram of $g$ in these bins is nearly flat.  We then take the counts in each bin as an estimate of the probability, compute the information and examine the dependence on the number of bins and the number of samples.  Following Refs \cite{strong+al_98,slonim+al_05}, we search for expected systematic dependencies, and extrapolate to the limit where the number of bins and samples both become large.    We can  obtain an upper bound on the information by assuming that the conditional distribution $P(g|x)$ is Gaussian, and we can obtain an approximation to the information by taking this Gaussian approximation through to the construction of $P_g(g)$; all these estimation procedures agree within error bars.

{\em Analysis of multiple genes.}  With simultaneous measurements of expression levels for multiple genes, we can estimate the information that they carry jointly.  The difficulty is that the space of expression levels is now much larger, but our number of samples is not.  Having calibrated the Gaussian approximation against more direct calculations for single genes (above), we can use this approximation in the multiple gene case, using Eq (\ref{jointPg}) directly in the integrals that define $I_{\{g_{\rm i}\}\rightarrow x}$.  We use a Monte Carlo method to evaluate these integrals numerically, and estimate errors by a bootstrap method. Importantly, if the signals that we observe are invertible linear combinations of the true signals---as might happen, for example, because of a small amount of crosstalk among the different imaging channels---then the invariance of the information to coordinate transformations tells us that this will not change our estimate.

\end{document}